\documentclass[lettersize,journal]{IEEEtran}
\usepackage{amsmath,amsfonts,amssymb}
\usepackage{algorithm}
\usepackage{algorithmic}

\usepackage{bm}
\usepackage{array}
\usepackage[caption=false,font=footnotesize,labelfont=rm,textfont=rm]{subfig}
\usepackage{textcomp}
\usepackage{stfloats}
\usepackage{url}
\usepackage{verbatim}
\usepackage{graphicx}
\usepackage{cite}
\usepackage{hyperref}
\hypersetup{hypertex=true,
	colorlinks=true,
	linkcolor=black,
	anchorcolor=black,
	citecolor=black}
\allowdisplaybreaks[2]
\makeatletter
\renewcommand{\maketag@@@}[1]{\hbox{\m@th\normalsize\normalfont#1}}%
\makeatother
\begin{document}

\title{Sum Rate Maximization for Movable Antenna Enhanced Multiuser Covert Communications}

\author{{Haobin Mao},~
	{Xiangyu Pi},~
	{Lipeng Zhu},~\IEEEmembership{Member,~IEEE,}
	{Zhenyu Xiao},~\IEEEmembership{Senior Member,~IEEE,}\\
	{Xiang-Gen Xia},~\IEEEmembership{Fellow,~IEEE,}
	and {Rui Zhang},~\IEEEmembership{Fellow,~IEEE}
	
	\thanks{H. Mao, X. Pi, and Z. Xiao are with the School of Electronic and Information Engineering and the State Key Laboratory of CNS/ATM, Beihang University, Beijing 100191, China (e-mail: maohaobin@buaa.edu.cn, pixiangyu@buaa.edu.cn, xiaozy@buaa.edu.cn). \textit{(Corresponding author: Lipeng Zhu and Zhenyu Xiao.)}} 
	\thanks{L. Zhu is with the Department of Electrical and Computer Engineering, National University of Singapore, Singapore 117583 (e-mail:zhulp@nus.edu.sg).}
	\thanks{X.-G. Xia is with the Department of Electrical and Computer Engineering, University of Delaware, Newark, DE 19716 USA (e-mail: xxia@ee.udel.edu).}
	\thanks{R. Zhang is with School of Science and Engineering, Shenzhen Research Institute of Big Data, The Chinese University of Hong Kong, Shenzhen, Guangdong 518172, China (e-mail: rzhang@cuhk.edu.cn). He is also with the Department of Electrical and Computer Engineering, National University of Singapore, Singapore 117583 (e-mail: elezhang@nus.edu.sg).}\vspace{-0.9cm}}


\maketitle

\begin{abstract}
In this letter, we propose to employ movable antenna (MA) to enhance covert communications with noise uncertainty, where the confidential data is transmitted from an MA-aided access point (AP) to multiple users with a warden attempting to detect the existence of the legal transmission. To maximize the sum rate of users under covertness constraint, we formulate an optimization problem to jointly design the transmit beamforming and the positions of MAs at the AP. To solve the formulated non-convex optimization problem, we develop a block successive upper-bound minimization (BSUM) based algorithm, where the proximal distance algorithm (PDA) and the successive convex approximation (SCA) are employed to optimize the transmit beamforming and the MAs' positions, respectively. Simulation results show that the proposed MAs-aided system can significantly increase the covert sum rate via antenna position optimization as compared to conventional systems with fixed-position antennas (FPAs).
\end{abstract}


\section{Introduction}
\IEEEPARstart{W}{ith} the increasing demand for massive and heterogeneous data transmission in the next-generation wireless network, multiple-input multiple-output (MIMO) communication has been recognized as a key enabling technology to achieve higher capacity and reliability. However, fixed-position antennas (FPAs) are usually adopted in conventional MIMO systems, which hinders the full utilization of the degrees of freedom (DoF) in the continuous spatial domain and thus limits the spatial multiplexing performance gain \cite{zhu2024modeling}. To overcome this limitation, movable antenna (MA) has been proposed to exploit the spatial DoF more effectively via flexible adjustment of the antenna position \cite{zhu2024movable,ning2024movable}, which is also known as fluid antenna system (FAS) \cite{zhu2024historical}. The great potentials of MAs have recently been demonstrated in various wireless systems, such as terrestrial transmissions \cite{zhu2024multiuser}, unmanned aerial vehicle (UAV) networks \cite{liu2024uav}, satellite communications \cite{zhu2024satellite}, non-orthogonal multiple access (NOMA) communications \cite{li2024noma}, and over-the-air computation systems \cite{li2024over}.


It is yet noteworthy that the inherent open nature of the wireless propagation environment renders secure transmissions vulnerable. Owing to the advantages of reconfiguring the wireless channel conditions, MAs can be employed to enlarge the difference between the received power gains in the desired and undesired directions to enhance physical layer security, which has recently gained increasing attention in the literature. The authors in \cite{hu2024secure} employed a linear MA array to enhance physical layer security against multiple eavesdroppers, where the transmit beamforming and MAs' positions were jointly designed. The authors in \cite{hu2024movable} minimized the secrecy outage probability by joint optimization of transmit beamforming and MAs' positions under imperfect channel state information of the eavesdroppers. The authors in \cite{cheng2024enabling} proposed an iteration algorithm to optimize the transmit beamformer and MAs' positions to maximize the secrecy rate in the presence of a legitimate user and an eavesdropper. 

To further enhance physical layer security, covert communication technologies have been proposed for transmitting private data against hostile detection. However, there is very limited work on MA-enabled covert communications. As a preliminary attempt, the authors in \cite{cheng2024covert} utilized a linear movable frequency diverse array to achieve the covert rate maximization of a single user via the joint optimization of antenna beamforming, frequencies, and positions. Different from this previous work, we propose to deploy a two-dimensional (2D) MA array at the access point (AP) to enhance multiuser covert communications with noise uncertainty in this letter. We aim to maximize the sum rate of users, subject to covertness constraint, by joint optimization of the transmit beamforming and positions of MAs at the AP. An efficient algorithm based on the block successive upper-bound minimization (BSUM) is developed to tackle the formulated highly non-convex optimization problem. Numerical results demonstrate that the proposed MA-enhanced covert communication design can achieve significantly higher performance gain in terms of sum rate than conventional systems with FPAs.

\section{System Model and Problem Formulation}
\subsection{System Model}
\begin{figure}[t]
	\setlength{\abovecaptionskip}{0cm}
	\begin{center}
		\includegraphics[width= 7.8 cm]{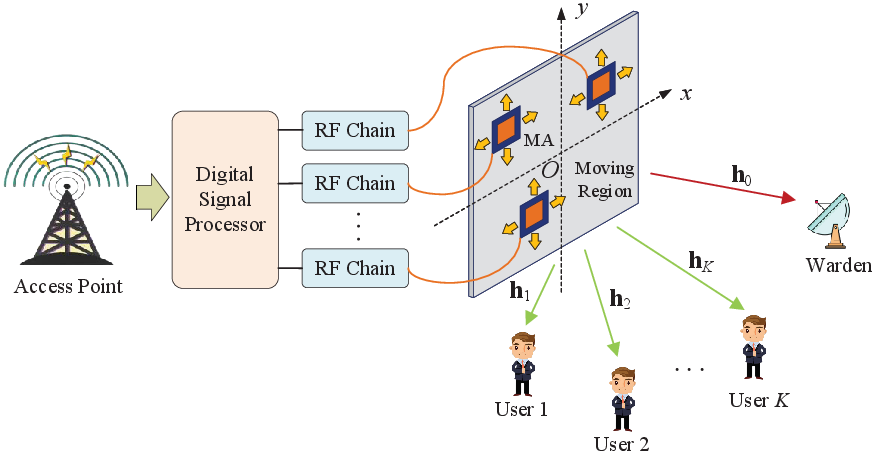}
		\caption{Illustration of the MAs enhanced covert communications.}
		\label{fig:Scenario}
	\end{center}
	\vspace{-2em}
\end{figure}

As shown in Fig. \ref{fig:Scenario}, we consider a multiuser covert communication system, where the AP equipped with $N$ MAs attempts to transmit confidential information to $K$ users while hiding the legitimate transmission from the hostile detection of a warden. We assume that the users and the warden both employ a single FPA for communication and detection, respectively. Denote the position of the $n$-th MA as $\mathbf{t}_n = \left[x_n,y_n\right]^{\mathsf{T}} \in \mathcal{M}$ for $1 \le n \le N$, where $\mathcal{M}$ is the given square moving region for MAs with the size of $A \times A$. Denote the set of coordinates of $N$ MAs by ${\mathbf{T}} = \{ \mathbf{t}_1,\mathbf{t}_2,...,\mathbf{t}_N \}$.
Let $L_k, 1 \le k \le K$, represent the numbers of receive channel paths from the AP to user $k$. The normalized direction for the $\ell$-th path from the AP to user $k$ is thus given by ${\bm{\rho }}_{k,\ell} = \left[\sin \theta_{k,\ell} \cos \phi_{k,\ell}, \cos \theta_{k,\ell} \right]^{\mathsf{T}}, 1 \le \ell \le L_k$, where $\theta_{k,\ell}$ and $\phi_{k,\ell}$ are the elevation and azimuth angles of departure (AoD), respectively. Then, the field-response vector of the receive channel paths between the $n$-th MA at the AP and user $k$ can be defined as ${\mathbf{f}}_k(\mathbf{t}_n) =  {\left[ {{{\rm{e}}^{{\rm{j}}\frac{{{\rm{2\pi }}}}{{\rm{\lambda }}}{\bf{t}}_n^{\mathsf{T}}{{\bm{\rho }}_{k,1}}}},{{\rm{e}}^{{\rm{j}}\frac{{{\rm{2\pi }}}}{{\rm{\lambda }}}{\bf{t}}_n^{\mathsf{T}}{{\bm{\rho }}_{k,2}}}},...,{{\rm{e}}^{{\rm{j}}\frac{{{\rm{2\pi }}}}{{\rm{\lambda }}}{\bf{t}}_n^{\mathsf{T}}{{\bm{\rho }}_{k,{L_k}}}}}} \right]^{\mathsf{T}}}$, where $\lambda$ is the carrier wavelength. Let ${{\mathbf{g }}_k} = {\left[g_{k,1},g_{k,2},...,g_{k,L_k} \right]}^{\mathsf{T}} \in \mathbb{C}^{L_k \times 1}$ be the path response vector between the AP and user $k$. We assume that the channels in the considered multiple-input single-output (MISO) system are quasi-static flat-fading. By adopting the field-response based channel model \cite{zhu2024modeling}, the channel vector between the AP and user $k$ is given by
\begin{equation}
{{\bf{h}}_k}({\mathbf{T}}) = {\mathbf{F}}_k^{\mathsf{H}}({\mathbf{T}}){{\mathbf{g}}_k} \in \mathbb{C}^{N \times 1}, 1 \le k \le K,
\end{equation}
where $\mathbf{F}_k({\mathbf{T}}) = \left[{\mathbf{f}}_k(\mathbf{t}_1), {\mathbf{f}}_k(\mathbf{t}_2),..., {\mathbf{f}}_k(\mathbf{t}_N) \right] \in \mathbb{C}^{L_k \times N}$. Similarly, the channel vector between the AP and the warden can be modelled as ${\bf{h}}_0({\mathbf{T}})$. In this letter, we assume that the channel state information of the warden can be obtained at the AP by adopting some advanced detection equipments, such as Ghostbuster \cite{chaman2018ghostbuster}.

By employing the linear beamformer at the AP, the receive signal-to-interference-plus-noise (SINR) of user $k$ is expressed as
\begin{equation}
\gamma_k = {\frac{{{{\left| {{\bf{h}}_k^{\mathsf{H}}({\mathbf{T}}){{\mathbf{w}}_k}} \right|}^2}}}{{\sum\nolimits_{i = 1,i \ne k}^K {{{\left| {{\bf{h}}_k^{\mathsf{H}}({\mathbf{T}}){{\mathbf{w}}_i}} \right|}^2}}  + {\sigma ^2}}}}, 1 \le k \le K,
\end{equation}
where ${{\mathbf{w}}_k}$ is the AP beamforming vector and $\sigma^2$ is the average power of the zero-mean additive white Gaussian noise (AWGN).


\subsection{Hostile Detection Performance}
Generally, the warden conducts a binary hypothesis testing to detect the presence of the legitimate transmission from the AP to users. Assuming that the warden employs the energy detection technique, the hypothesis testing problem can be modelled as 
\begin{equation}
\left\{ {\begin{array}{*{20}{l}}
	{{{\mathcal H}_0}:{\mathcal T}_w = \sigma_w^2},\\
	{{{\mathcal H}_1}:{\mathcal T}_w = {\sum\nolimits_{k = 1}^K {{{\left| {{\bf{h}}_0^{\mathsf{H}}{{\mathbf{w}}_k}} \right|}^2}}}  + \sigma_w^2},
	\end{array}} \right.
\end{equation}
where ${\mathcal H}_0$ represents the null hypothesis that the AP keeps silent, and ${\mathcal H}_1$ denotes the alternative hypothesis. 
${\mathcal T}_w$ is the average power of the received signal at the warden. $\sigma_w^2$ is the average power of the AWGN at the warden. We assume that the exact noise power at the warden follows a log-uniform distribution with the probability density function (PDF) given by
\begin{equation}
f_{\sigma_{w}^2}\left( x \right) = \left\{ {\begin{array}{*{20}{l}}
		\frac{1}{2 {\rm{ln}}(\tau) x},&\frac{1}{\tau}\hat{\sigma}_w^2 \le x \le {\tau}\hat{\sigma}_w^2,\\
		0,&\rm{otherwise},
\end{array}}\right.
\end{equation}
where $\tau \ge 1$ measures the uncertainty of the noise power and $\hat{\sigma}_w^2$ is the nominal noise power at the warden \cite{zhou2019trajectory}. Note that due to the uncertainty of the environmental noise power, the warden may make a wrong decision on whether the AP is transmitting. According to \cite{zhou2019trajectory}, the optimal detection threshold can be set as $\delta_{\rm th}^* = \sum\nolimits_{k = 1}^K {{{\left| {{\bf{h}}_0^{\mathsf{H}}({\mathbf{T}}){{\mathbf{w}}_k}} \right|}^2}} + \frac{1}{\tau}\hat{\sigma}_w^2$ and the minimum total detection error probability is given by
\begin{equation}
\xi^* = 1 - \frac{1}{2\rm{ln}(\tau)}{\rm{ln}}\left( 1 + \frac{\sum\nolimits_{k = 1}^K {{{\left| {{\bf{h}}_0^{\mathsf{H}}({\mathbf{T}}){{\mathbf{w}}_k}} \right|}^2}} }{\hat{\sigma}_w^2/{\tau}}\right).
\end{equation}

To achieve high-level security of the considered system, we need to ensure the minimum total detection error probability to be close to $1$, i.e., $\xi^* \ge 1 - \varepsilon$, which is known as the covertness constraint, where $\varepsilon$ is a small positive value to measure the system covertness. As such, by defining $p_{\rm{th}} = \hat{\sigma}_w^2 ({\tau}^{2\varepsilon} - 1)/\tau$, the covertness constraint can be equivalently transformed to
\begin{equation} \label{covertness}
\sum\nolimits_{k = 1}^K {{{\left| {{\bf{h}}_0^{\mathsf{H}}({\mathbf{T}}){{\mathbf{w}}_k}} \right|}^2}} \le p_{\rm{th}}.
\end{equation}
\vspace{-0.5 cm}
\subsection{Problem Formulation}
In this letter, we aim to maximize the sum rate of the users by jointly designing the transmit beamformer and the positions of MAs at the AP under the covertness constraint. Denote $\mathbf{W} = \left[ {\mathbf{w}}_1,{\mathbf{w}}_2,...,{\mathbf{w}}_K\right] \in \mathbb{C}^{N \times K}$ as the beamformer at the AP. Accordingly, the optimization problem is formulated as 
\begin{subequations}\label{opti}
	\begin{align}
	\max_{\left\{\mathbf{W},\mathbf{T} \right\}}& \sum\nolimits_{k = 1}^{K} \log_2 \left(1 + \gamma_k \right)\\
	\text{s.t.} \ & \mathcal{C}_{\rm 1}: \sum\nolimits_{k = 1}^{K}  \left\| \mathbf{w}_k\right\| ^2 \le p_{\rm max} \label{opti:sub1},\\
	&\mathcal{C}_{\rm 2}: \mathbf{t}_n \in \mathcal{M}, \left\| {\bf{t}}_n - {\bf{t}}_{n'} \right\|_2 \ge {d_{\min }}, \nonumber\\
	& \ \ \ \ \ 1 \le n,n' \le N, n \ne n', \label{opti:sub2}\\
	&\mathcal{C}_{\rm 3}: (\ref{covertness}),
	\end{align}
\end{subequations}
where (\ref{opti:sub1}) guarantees that the transmit power at the AP does not exceed the maximum value. (\ref{opti:sub2}) limits the antenna moving region and the distance between any two MAs. It is noted that problem (\ref{opti}) is challenging to solve due to the highly non-concave objective function and non-convex constraints. Meanwhile, $\bf{W}$ and $\bf{T}$ are highly coupled, which makes the problem more intricate to solve. Therefore, we develop a BSUM based algorithm to obtain a suboptimal solution for problem (\ref{opti}) in the next section.

\section{Proposed Solution}
By employing the idea of the weighted minimum mean square error (WMMSE) method \cite{shi2011mmse}, we equivalently transform the original problem to the following problem,
\begin{equation}
\min_{\left\{\bm{\varphi},\bm{\beta},\mathbf{W},\mathbf{T} \right\}} f(\bm{\varphi}, \bm{\beta}, \mathbf{W},\mathbf{T}) \ \ \ \ \text{s.t.} \ \mathcal{C}_{\rm 1},\mathcal{C}_{\rm 2},\mathcal{C}_{\rm 3},
\end{equation}
where $f\left( {\bm{\varphi},\bm{\beta},{\mathbf{W,T}}} \right) = \sum\nolimits_{k = 1}^K {{\beta _k}{g_k}\left( {\varphi_k,{\mathbf{W,T}}} \right) - \log \left( {{\beta _k}} \right)}$, $\bm{\varphi} = \left[ \varphi_1,\varphi_2,\cdots,\varphi_K \right] \in \mathbb{C}^{1 \times K} $ and $\bm{\beta} = \left[ \beta_1,\beta_2,\cdots,\beta_K \right] \in\mathbb{R}^{1 \times K}_{++} $ are two auxiliary vector variables, and ${g_k}\left( \varphi_k,\mathbf{W,T} \right) =  \left| \varphi_k \right|^2 \left(\sum\nolimits_{i = 1}^K \left| {\bf{h}}_k^ \mathsf{H} {\mathbf{w}}_i \right|^2  + {\sigma ^2} \right) - 2 \Re \{ \varphi_k^* {\bf{h}}_k^ \mathsf{H} {\mathbf{w}}_k \} + 1$. $\Re\{x\}$ denotes the real part of complex number $x$.

After the above transformation, the original problem is more tractable and can be efficiently solved by applying the BSUM method, which mainly involves the following variable updating steps in say, the $(m+1)$-th iteration, $m\geq 1$.

\begin{subequations}\small
	\setlength\abovedisplayskip{0cm}
	\begin{align}
	&\bm{\varphi}^{(m+1)} = \arg \min_{\bm{\varphi} \in \mathbb{C}^{1 \times K}} f(\bm{\varphi}, \bm{\beta}^{(m)}, \mathbf{W}^{(m)},\mathbf{T}^{(m)}) \label{opti1:sub1},\\
	&\bm{\beta}^{(m+1)} = \arg \min_{\bm{\beta} \in \mathbb{R}^{1 \times K}_{++}} f(\bm{\varphi}^{(m+1)}, \bm{\beta}, \mathbf{W}^{(m)},\mathbf{T}^{(m)}) \label{opti1:sub2}, \\
	&\mathbf{W}^{(m+1)} = \arg \min_{\mathbf{W} \in \mathcal{C}_1 \cap \mathcal{C}_3} f(\bm{\varphi}^{(m+1)}, \bm{\beta}^{(m+1)}, \mathbf{W},\mathbf{T}^{(m)}) \label{opti1:sub3}, \\
	&\mathbf{T}^{(m+1)} = \arg \min_{\mathbf{T} \in \mathcal{C}_2 \cap \mathcal{C}_3} f(\bm{\varphi}^{(m+1)}, \bm{\beta}^{(m+1)}, \mathbf{W}^{(m+1)},\mathbf{T}) \label{opti1:sub4}, 
	\end{align}
\end{subequations}
where $\bm{\varphi}^{(m)}, \bm{\beta}^{(m)}, \mathbf{W}^{(m)}$, and $\mathbf{T}^{(m)}$ are the solutions obtained in the $m$-th iteration. It is worth noting that the closed-form solutions to problems (\ref{opti1:sub1}) and (\ref{opti1:sub2}) can be obtained as \cite{shi2011mmse}:

\begin{equation}\label{phi1}
\varphi_k^{(m+1)} = \frac{{\bf{h}}_k^ \mathsf{H}(\mathbf{T}^{(m)})\mathbf{w}_k^{(m)}}{\sum\nolimits_{i = 1}^K \left| {\bf{h}}_k^ \mathsf{H}(\mathbf{T}^{(m)}) {\mathbf{w}}_i^{(m)} \right|^2  + {\sigma ^2}},1 \le k \le K,
\end{equation}
and
\begin{equation}\label{beta1}
\beta_k^{(m+1)} = \left( 1 - ({\varphi}_k^{(m+1)})^*{\bf{h}}_k^ \mathsf{H}(\mathbf{T}^{(m)})\mathbf{w}_k^{(m)} \right)^{-1},1 \le k \le K.
\end{equation}

However, problems (\ref{opti1:sub3}) and (\ref{opti1:sub4}) are still highly non-convex and difficult to solve. Therefore, we propose to leverage the proximal distance algorithm (PDA) and SCA based method to tackle the two problems, respectively.

\subsection{Transmit Beamforming Optimization}
Given $\bm{\varphi}^{(m+1)}$, $\bm{\beta}^{(m+1)}$, and $\mathbf{T}^{(m)}$, the optimization problem (\ref{opti1:sub3}) for transmit beamforming can be formulated as
\begin{subequations}\label{beamforming1}
	\begin{align}
		\min_{\left\{\mathbf{W}\right\}}& \sum\nolimits_{k = 1}^K {\bf{w}}_k^\mathsf{H}{\bf{A}}_k{\bf{w}}_k  - \sum\nolimits_{k = 1}^K {\Re} \left\{ {2{\bf{b}}_k^{\mathsf{H}}{{\bf{w}}_k}} \right\} \\
 		\text{s.t.} \ & \mathcal{C}_{\rm 1},\mathcal{C}_{\rm 3}: \sum\nolimits_{k = 1}^K \mathbf{w}_k^{\mathsf{H}} \mathbf{H}_0\mathbf{w}_k \le p_{\rm{th}},
	\end{align}
\end{subequations}
where ${\bf{A}}_k = \sum\nolimits_{k = 1}^K {\beta _k^{(m + 1)}{{\left| {\varphi_k^{(m + 1)}} \right|}^2}{{\bf{h}}_k}({\mathbf{T}^{(m)}}){\bf{h}}_k^{\mathsf{H}}\left( {\mathbf{T}^{(m)}} \right)}$, ${{\bf{b}}_k} = \beta_k^{(m + 1)} \varphi_k^{(m + 1)}{{\bf{h}}_k}(\mathbf{T}^{(m)})$, and $\mathbf{H}_0 = {{\bf{h}}_0}({\mathbf{T}^{(m)}}){\bf{h}}_0^{\mathsf{H}}({\mathbf{T}^{(m)}})$. By utilizing the PDA, problem (\ref{beamforming1}) can be recast to the following problem \cite{zhang2024pda}:
\begin{equation} \label{reformulate}
\begin{aligned}
\min_{\left\{\mathbf{W}\right\}} &\sum\nolimits_{k = 1}^K {\bf{w}}_k^\mathsf{H}{\bf{A}}_k{\bf{w}}_k  - \sum\nolimits_{k = 1}^K {\Re} \left\{ {2{\bf{b}}_k^{\mathsf{H}}{{\bf{w}}_k}} \right\} \\
&+ \varrho \cdot {\rm{dist}}^2({\bf{W}},\mathcal{C}_1) + \varrho \cdot {\rm{dist}^2}({\bf{W}},\mathcal{C}_3),
\end{aligned}
\end{equation}
where $\varrho > 0$ is the penalty factor and ${\rm{dist}}({\bf{x}},\mathcal{X}) = \min_{{\bf{y}} \in \mathcal{X}} \left\| \bf{x} - \bf{y} \right\|_2$ denotes the distance function from point $\bf{x}$ to set $\mathcal{X}$. Note that problem (\ref{reformulate}) is equivalent to problem (\ref{beamforming1}) as $\varrho$ approaches infinity.

Then, we substitute the distance functions with their upper bounds, which are given by
\begin{equation}
{\rm{dist}}({\bf{W}},\mathcal{C}_1) \le \left\| {\bf{W}}-\widetilde{{\bf{W}}}_1 \right\|_2,{\rm{dist}}({\bf{W}},\mathcal{C}_3) \le \left\| {\bf{W}}-\widetilde{{\bf{W}}}_2 \right\|_2,
\end{equation}
where $\widetilde{{\bf{W}}}_1 = \Pi_{\mathcal{C}_1}({\bf{W}}^{(m)})$ and $\widetilde{{\bf{W}}}_2 = \Pi_{\mathcal{C}_3}({\bf{W}}^{(m)})$ with $\Pi_{\mathcal{X}}({\bf{x}}) = \arg\min_{{\bf{y}} \in \mathcal{X}} \left\| \bf{x} - \bf{y} \right\|_2$ representing the projection of $\mathbf{x}$ onto set $\mathcal{X}$, which can be obtained by employing the Lagrange multiplier method for any given $\mathbf{W}^{(m)}$ as
\begin{equation}
[{{\bf{\widetilde {\bf{W}}}}_{1}}]_{:,k} = \left\{ {\begin{array}{*{20}{l}}
	{{{\bf{w}}_k^{(m)}},{\rm{if}}\sum\nolimits_{k = 1}^K \left\| \mathbf{w}_k^{(m)}\right\| ^2 \le p_{\rm max}},\\
	{\sqrt {{p_{\max }}} \frac{{{{\bf{w}}_k^{(m)}}}}{{\sum\nolimits_{k = 1}^K ({{\bf{w}}_k^{(m)})^{\rm{H}}{{\bf{w}}_k^{(m)}}} }},{\rm{otherwise}}},
	\end{array}} \right.
\end{equation}
\begin{equation}
[{{\bf{\widetilde {\bf{W}}}}_{2}}]_{:,k} = \left\{ {\begin{array}{*{20}{l}}
	{{{\bf{w}}_k}^{(m)},{\rm{if}}\sum\nolimits_{k = 1}^K {{{\left| {{\bf{h}}_0^{\mathsf{H}}(\mathbf{T}^{(m)}){{\mathbf{w}}_k^{(m)}}} \right|}^2}} \le p_{\rm{th}}},\\
	{{{\left( {{{\bf{I}}_N} + \varsigma {{\bf{h}}_0(\mathbf{T}^{(m)})}{\bf{h}}_0^{\mathsf{H}}}(\mathbf{T}^{(m)}) \right)}^{ - 1}}{{\bf{w}}_k^{(m)}},{\rm{otherwise}}},
	\end{array}} \right.
\end{equation}
where $\varsigma \ge 0$ can be obtained by one-dimensional bisection search until $\sum\nolimits_{k = 1}^K {{{\left| {\bf{h}}_0^{\mathsf{H}}(\mathbf{T}^{(m)}) \left[{{\bf{\widetilde {\bf{W}}}}_{2}}\right]_{:,k} \right|}^2}} = p_{\rm{th}}$ is satisfied. $\left[ \mathbf{X} \right]_{:,j}$ represents the $j$-th column of matrix $\mathbf{X}$. Hereto, problem (\ref{reformulate}) can be rewritten as
\begin{equation} \label{beamforming} 
\begin{aligned}
\min_{\left\{\mathbf{W}\right\}} &\sum\nolimits_{k = 1}^K {\bf{w}}_k^\mathsf{H}{\bf{A}}_k{\bf{w}}_k  - \sum\nolimits_{k = 1}^K {\Re} \left\{ {2{\bf{b}}_k^{\mathsf{H}}{{\bf{w}}_k}} \right\} \\
&+ \varrho \cdot \left\| {\bf{W}}-\widetilde{{\bf{W}}}_1 \right\|_2^2 + \varrho \cdot \left\| {\bf{W}}-\widetilde{{\bf{W}}}_2 \right\|_2^2,
\end{aligned}
\end{equation}
which is an unconstrained quadratic problem and can be efficiently solved by the Lagrange multiplier method. The optimal solution to problem (\ref{beamforming}) is given by
\begin{equation} \label{beamsolution}
{\bf{w}}_k^{(m + 1)} = {\left( {{\bf{A}}_k + 2 \varrho {{\bf{I}}_N}} \right)^{ - 1}}\left( {{{\bf{b}}_k} + \varrho [{{\bf{\widetilde {\bf{W}}}}_{1}}]_{:,k} + \varrho [{{\bf{\widetilde {\bf{W}}}}_{2}}]_{:,k}} \right).
\end{equation}
By gradually increasing the value of $\varrho$, the proposed PDA based method for transmit beamforming can converge to a suboptimal solution to problem (\ref{beamforming1}), which is summarized in Algorithm \ref{alg1}. Note that we need to initialize the penalty factor $\varrho$ to be sufficiently small to obtain a satisfactory solution to problem (\ref{beamforming1}).
\begin{algorithm}[t]
	\caption{PDA-based algorithm for beamforming}
	\label{alg1}
	\begin{algorithmic}[1]
		\REQUIRE
		$\bm{\varphi}^{(m+1)}, \bm{\beta}^{(m+1)}, \mathbf{W}^{(m)},\mathbf{T}^{(m)}$.
		\ENSURE
		$\mathbf{W}^{(m+1)}$.
		\STATE
		Initialize $\varrho > 0$, $\kappa > 1$, the iteration index $s = 1$, and $\mathbf{W}^{(s)} = \mathbf{W}^{(m)}$.
		\REPEAT	
		\STATE Update $\mathbf{W}^{(s+1)}$ according to (\ref{beamsolution}).
		\STATE $\mathbf{W}^{(m+1)} \leftarrow \mathbf{W}^{(s+1)}$, $\varrho \leftarrow \kappa \varrho$, and $s \leftarrow s+1$.
		\UNTIL The decrease of the objective of problem (\ref{beamforming}) is no larger than $10^{-3}$ or the maximum iteration number $S_{\max}$ is reached.
	\end{algorithmic}
\end{algorithm}
\vspace{-0.2cm}
\subsection{Antenna Position Optimization}
Note that problem (\ref{opti1:sub4}) is difficult to solve because the MAs' positions are coupled in the objective function. In this letter, we resort to optimizing the position of each antenna in a sequential manner with the positions of all the other antennas being fixed. Therefore, given the antenna positions $\{{\bf{t}}_{n'}, 1 \le n' \le N, n' \neq n\}$, $\bm{\varphi}^{(m+1)}$, $\bm{\beta}^{(m+1)}$, and $\mathbf{W}^{(m+1)}$, the subproblem of optimizing the $n$-th antenna position can be reduced to 
\begin{subequations}\label{position1}
	\begin{align}
	\min_{\left\{\mathbf{t}_{n} \right\}} & \sum\nolimits_{k=1}^K \beta_k^{(m+1)} \Psi_k(\mathbf{t}_{n})\\
	\text{s.t.} \ &\mathcal{C}_{\rm 2},\mathcal{C}_{\rm 3}:\sum\nolimits_{k = 1}^K {{{\left| {({\mathbf{w}}_k^{(m+1)})}^{\mathsf{H}} {\bf{h}}_0({{\bf{t}}_{n}}) \right|}^2}} \le p_{\rm{th}}.
	\end{align}
\end{subequations}
where $\Psi(\mathbf{t}_{n}) =  \left| \varphi_k^{(m+1)} \right|^2 \sum\nolimits_{i = 1}^K \left| {{({\mathbf{w}}_i^{(m+1)}})^{\mathsf{H}}{\bf{h}}_k(\mathbf{t}_{n})} \right|^2 - 2 \Re \{ \varphi_k^{(m+1)} {({\mathbf{w}}_k^{(m+1)}})^{\mathsf{H}} {\bf{h}}_k(\mathbf{t}_{n})  \}$. 

It is noted that problem (\ref{position1}) is still non-convex and challenging to solve. We utilize the SCA method to relax problem (\ref{position1}) as a convex one and then obtain a suboptimal solution. By defining $\alpha_{k,i}^{n,n',\ell,\ell'} =  w_{i,n}^{(m+1)} (w_{i,n'}^{(m+1)})^{*}g_{k,\ell}^{*} g_{k,\ell '}$ with $w_{i,n}^{(m+1)}$ denoting the $n$-th element of ${\mathbf{w}}_i^{(m+1)}$, we rewrite the term $\left| {{({\mathbf{w}}_i^{(m+1)}})^{\mathsf{H}}{\bf{h}}_k({{\bf{t}}_{n}})} \right|^2$ as (\ref{formula1}) shown at the bottom of the next page, where the constant $\rm{C}_1$ is given by
\begin{figure*}[hb]
	\centering
	\vspace*{-1em}
	\hrulefill
	\vspace*{-5pt} 
	\begin{equation} \label{formula1}
	\begin{aligned}
	\left| {{({\mathbf{w}}_i^{(m+1)}})^{\mathsf{H}}{\bf{h}}_k({{\bf{t}}_{n}})} \right|^2 &= {\rm{C_1}} + \sum\nolimits_{\ell = 1}^{L_k}\sum\nolimits_{\ell ' = 1}^{L_k} \mid \alpha_{k,i}^{n,n,\ell,\ell'}\mid \cdot \cos \left( {\frac{{2\pi }}{\lambda }} {\bf{t}}_n^{\mathsf{T}} \left( {{\bm{\rho }}_{k,\ell}} - {{\bm{\rho }}_{k,\ell'}}\right) + \angle \alpha_{k,i}^{n,n,\ell,\ell'} \right) \\
	&+ 2\sum\nolimits_{\substack{n'=1,n' \neq n} }^{N} \sum\nolimits_{\ell = 1}^{L_k}\sum\nolimits_{\ell ' = 1}^{L_k} \mid \alpha_{k,i}^{n,n',\ell,\ell'}\mid \cdot \cos \left( {\frac{{2\pi }}{\lambda }} \left({\bf{t}}_n^{\mathsf{T}}{{\bm{\rho }}_{k,\ell}} - {\bf{t}}_{n'}^{\mathsf{T}}{{\bm{\rho }}_{k,\ell'}} \right) + \angle{\alpha_{k,i}^{n,n',\ell,\ell'}}\right)
	\end{aligned}
	\end{equation}
	\vspace*{-1em} 
	\hrulefill
	\vspace*{-1pt}
\end{figure*}
\begin{equation}
\begin{aligned}
{\rm{C_1}} &= \sum\nolimits_{\substack{n'=1,n' \neq n} }^{N}\sum\nolimits_{\substack{n''=1,n'' \neq n }}^{N} \sum\nolimits_{\ell = 1}^{L_k}\sum\nolimits_{\ell ' = 1}^{L_k} \mid \alpha_{k,i}^{n',n'',\ell,\ell'}\mid \\
&\cdot \cos \left( {\frac{{2\pi }}{\lambda }} \left({\bf{t}}_{n'}^{\mathsf{T}}{{\bm{\rho }}_{k,\ell}} - {\bf{t}}_{n''}^{\mathsf{T}}{{\bm{\rho }}_{k,\ell'}} \right) + \angle{\alpha_{k,i}^{n',n'',\ell,\ell'}} \right).
\end{aligned}
\end{equation}

 Then, we construct a convex upper-bound function of $\left| {{({\mathbf{w}}_i^{(m+1)}})^{\mathsf{H}}{\bf{h}}_k({{\bf{t}}_{n}})} \right|^2$ by employing the second-order Taylor expansion \cite{zhu2024satellite}, which is given by (\ref{formula2}) shown at the bottom of the next page with
 \begin{figure*}[!hb]
 	\centering
 	\begin{equation} \label{formula2}
 	\begin{aligned}
 	&\left| {{({\mathbf{w}}_i^{(m+1)}})^{\mathsf{H}}{\bf{h}}_k({{\bf{t}}_{n}})} \right|^2 \le {\rm{C_1}} + \sum\nolimits_{\ell = 1}^{L_k}\sum\nolimits_{\ell ' = 1}^{L_k} \mid \alpha_{k,i}^{n,n,\ell,\ell'}\mid \cdot \psi^{\rm ub}\left({{\bf{t}}_{n}} \mid {{\bf{t}}_{n}^{(m)}}, {{\bm{\rho }}_{k,\ell}} - {{\bm{\rho }}_{k,\ell'}}, \angle\alpha_{k,i}^{n,n,\ell,\ell'} \right) \\
 	&+ 2\sum\nolimits_{\substack{n'=1,n' \neq n} }^{N} \sum\nolimits_{\ell = 1}^{L_k}\sum\nolimits_{\ell ' = 1}^{L_k} \mid \alpha_{k,i}^{n,n',\ell,\ell'}\mid \cdot \psi^{\rm ub}\left({{\bf{t}}_{n}} \mid {{\bf{t}}_{n}^{(m)}}, {{\bm{\rho }}_{k,\ell}}, \angle\alpha_{k,i}^{n,n',\ell,\ell'}- {\frac{{2\pi }}{\lambda }}{{\bf{t}}_{n'}^{\mathsf{T}}}{{\bm{\rho }}_{k,\ell'}} \right) \triangleq  \zeta_{1}^{i,k} \left( {{\bf{t}}_{n}}\right).
 	\end{aligned}
 	\end{equation}
 	\vspace*{-1em} 
 	\hrulefill
 	\vspace*{-1pt}
 \end{figure*}

 \begin{equation}
 \setlength\abovedisplayskip{0cm}
 \begin{aligned}
 &\psi^{\rm ub}\left( {{\bf{t}}_{n}} \mid {{\bf{t}}_{n}^{(m)}}, {\bm{\rho }}, \alpha \right) = \cos \left( \frac{{2\pi}}{\lambda}{\bm{\rho }}^{\mathsf{T}}{{\bf{t}}_{n}^{(m)}}+ \angle \alpha \right)\\
 & - \frac{{2\pi }}{\lambda}\sin \left(\frac{{2\pi }}{\lambda}{\bm{\rho }}^{\mathsf{T}}{{\bf{t}}_{n}^{(m)}} + \angle \alpha \right){\bm{\rho }}^{\mathsf{T}}\left({{\bf{t}}_{n}} - {{\bf{t}}_{n}^{(m)}}\right)\\
 &+\frac{2\pi^2}{\lambda^2}\left({\bm{\rho }}^{\mathsf{T}}\left({{\bf{t}}_{n}} - {{\bf{t}}_{n}^{(m)}}\right)\right)^2.
 \end{aligned}
 \end{equation}

Similarly, by defining $\vartheta_k^{n,l} = w_{k,n}^{(m+1)} (\varphi_k^{(m+1)})^{*} g_{k,\ell}^{*}$, we rewrite the term $\Re \{ \varphi_k^{(m+1)} {({\mathbf{w}}_k^{(m+1)}})^{\mathsf{H}} {\bf{h}}_k({{\bf{t}}_{n}})  \}$ as
\begin{equation} \label{formula3}
\begin{aligned}
&\Re \{ \varphi_k^{(m+1)} {({\mathbf{w}}_k^{(m+1)}})^{\mathsf{H}} {\bf{h}}_k({{\bf{t}}_{n}})  \}\\
& = {\rm{C_2}} + \sum\nolimits_{\ell = 1}^{L_k}\mid \vartheta_{k}^{n,\ell}\mid \cdot \cos \left( {\frac{{2\pi }}{\lambda }} {\bf{t}}_n^{\mathsf{T}}{{\bm{\rho }}_{k,\ell}} + \angle \vartheta_{k}^{n,\ell} \right),
\end{aligned}
\end{equation}
where
\begin{equation}
{\rm{C_2}} = \sum\limits_{\substack{n'=1,n' \neq n}}^{N}\sum\nolimits_{\ell = 1}^{L_k}\mid \vartheta_{k}^{n',\ell}\mid \cdot \cos \left( {\frac{{2\pi }}{\lambda }} {\bf{t}}_{n'}^{\mathsf{T}}{{\bm{\rho }}_{k,\ell}} + \angle \vartheta_{k}^{n',\ell} \right)
\end{equation} is a constant.

By adopting the same idea of \cite{zhu2024satellite}, we construct a concave lower-bound of $\Re \{ \varphi_k^{(m+1)} {({\mathbf{w}}_k^{(m+1)}})^{\mathsf{H}}  {\bf{h}}_k({{\bf{t}}_{n}})  \}$ as
\begin{equation} \label{formula5}
\begin{aligned}
&\Re \{ \varphi_k^{(m+1)} {({\mathbf{w}}_k^{(m+1)}})^{\mathsf{H}} {\bf{h}}_k({{\bf{t}}_{n}})  \} \ge {\rm{C_2}}\\
& + \sum\nolimits_{\ell = 1}^{L_k}\mid \vartheta_{k}^{n,\ell}\mid \cdot \psi^{\rm lb}\left( {{\bf{t}}_{n}} \mid {{\bf{t}}_{n}^{(m)}},{{\bm{\rho }}_{k,\ell}}, {\vartheta}_{k}^{n,\ell} \right) \triangleq  \zeta_2^k \left( {{\bf{t}}_{n}}\right),
\end{aligned}
\end{equation}
where 
\begin{equation}
\begin{aligned}
&\psi^{\rm lb}\left(  {{\bf{t}}_{n}} \mid {{\bf{t}}_{n}^{(m)}},{\bm{\rho }}, {\vartheta} \right) = \cos \left( \frac{{2\pi}}{\lambda}{\bm{\rho }}^{\mathsf{T}}{{\bf{t}}_{n}^{(m)}}+ \angle \vartheta \right)\\
& - \frac{{2\pi }}{\lambda}\sin \left(\frac{{2\pi }}{\lambda}{\bm{\rho }}^{\mathsf{T}}{{\bf{t}}_{n}^{(m)}} + \angle \vartheta \right){\bm{\rho }}^{\mathsf{T}}\left({{\bf{t}}_{n}} - {{\bf{t}}_{n}^{(m)}}\right)\\
&-\frac{2\pi^2}{\lambda^2}\left({\bm{\rho }}^{\mathsf{T}}\left({{\bf{t}}_{n}} - {{\bf{t}}_{n}^{(m)}}\right)\right)^2.
\end{aligned}
\end{equation}

To handle the non-convex constraint (\ref{opti:sub2}), we use
\begin{equation}
\left\| {{{\bf{t}}_n} - {{\bf{t}}_{n'}}} \right\|_2 \ge \frac{{{{\left( {{\bf{t}}_n^{(m)} - {{\bf{t}}_{n'}}} \right)}^{\mathsf{T}}}\left( {{{\bf{t}}_n} - {{\bf{t}}_{n'}}} \right)}}{{\left\| {{\bf{t}}_n^{(m)} - {{\bf{t}}_{n'}}} \right\|}_2} \triangleq d_{n,n'}^{\rm{lb}}.
\end{equation}

To address the non-convexity of $\mathcal{C}_3$, we derive an upper-bound of  $\left| {{({\mathbf{w}}_k^{(m+1)}})^{\mathsf{H}}{\bf{h}}_0({{\bf{t}}_{n}})} \right|^2$ as $\zeta_3^k \left( {{\bf{t}}_{n}}\right)$ similar to (\ref{formula2}).

Thus, problem (\ref{position1}) can be relaxed as
\begin{subequations} \label{position2}
	\begin{align}
	\min_{\left\{\mathbf{t}_{n} \right\}} & \sum\limits_{k=1}^K \beta_k^{(m+1)} \left( \left| \varphi_k^{(m+1)} \right|^2 \sum\limits_{i = 1}^K \zeta_{1}^{i,k} \left( {{\bf{t}}_{n}}\right) - 2\zeta_2^k \left( {{\bf{t}}_{n}}\right)  \right)\\
	\text{s.t.} \ &\mathcal{C}'_{\rm 2}: \mathbf{t}_n \in \mathcal{M}, d_{n,n'}^{\rm{lb}} \ge {d_{\min }}, 1 \le n' \le N, n' \ne n, \\
	&\mathcal{C}'_{\rm 3}:\sum\nolimits_{k = 1}^K \zeta_3^k\left( {{\bf{t}}_{n}}\right) \le p_{\rm{th}},
	\end{align}
\end{subequations}
which is a convex problem and can be efficiently solved by the existing optimization tools, such as CVX \cite{cvx}.

We summarize the above BSUM-based algorithm for joint transmit beamforming and antenna position optimization in Algorithm \ref{alg2}, where the beamformer at the AP is initialized by the zero-forcing (ZF) method with the covertness constraint satisfied by reducing the transmit power, and the positions of MAs are initialized randomly following uniform distribution with the minimum distance between adjacent antennas larger than $d_{\rm min}$. It is noted that the objective value in each iteration can be guaranteed to be non-decreasing and is upper-bounded by a finite value. Therefore, the objective value is guaranteed to converge. The computational complexity of the proposed solution mainly lies in line 5 and can be shown to be in the order of $\mathcal{O}\left(M_{\rm{outer}}S_{\rm{inner}}N^{3} \right)$, where $M_{\rm{outer}}$ and $S_{\rm{inner}}$ are the iteration number of the outer loop and that of the loop in Algorithm \ref{alg1}, respectively.

\begin{algorithm}[!tb]
	\caption{BSUM-based algorithm for solving problem (\ref{opti})}
	\label{alg2}
	\begin{algorithmic}[1]
		\REQUIRE
		$N,K,L,\{\theta_{k,l}\},\{\phi_{k,l}\},\{\theta_{0,l}\},\{\phi_{0,l}\},{{\mathbf{\Sigma}}_k},{{\mathbf{\Sigma}}_0},p_{{\rm{max}}},$\\$\varepsilon,\lambda,\sigma^2,\hat{\sigma}_w^2,\mathcal{M},\mathbf{W}^{(1)},\mathbf{T}^{(1)}$.
		\ENSURE
		$\mathbf{W},\mathbf{T}$.
		\STATE
		Initialize the iteration index $m = 1$.
		\REPEAT
		\STATE Update $\bm{\varphi}^{(m+1)}$ according to (\ref{phi1}).
		\STATE Update $\bm{\beta}^{(m+1)}$ according to (\ref{beta1}).	
		\STATE Update $\mathbf{W}^{(m+1)}$ by using Algorithm \ref{alg1}.
		\FOR{$n=1$ to $N$}
		\STATE Update ${{\bf{t}}_{n}^{(m+1)}}$ by solving problem (\ref{position2}).
		\ENDFOR
		\STATE  $m \leftarrow m+1$.
		\UNTIL The increase of the sum rate is no larger than $10^{-3}$ or the maximum iteration number $M_{\max}$ is attained.
	\end{algorithmic}
\end{algorithm}
\vspace{-0.25cm}
\section{Numerical Results}
\begin{figure*}[t]
	\vspace{-2em}
	\setlength{\abovecaptionskip}{0.1cm}
	\centering
	\subfloat[Sum rate versus maximum transmit power.\label{fig:figure1}]{\includegraphics[width= 4 cm]{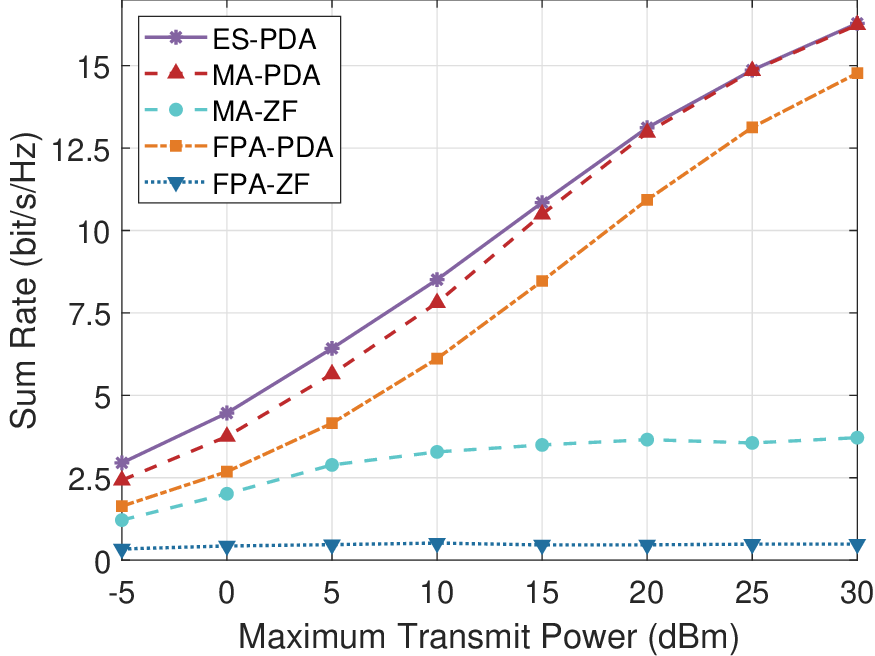}}%
	\hfil
	\subfloat[Sum rate versus numbers of antenna.\label{fig:figure2}]{\includegraphics[width= 4 cm]{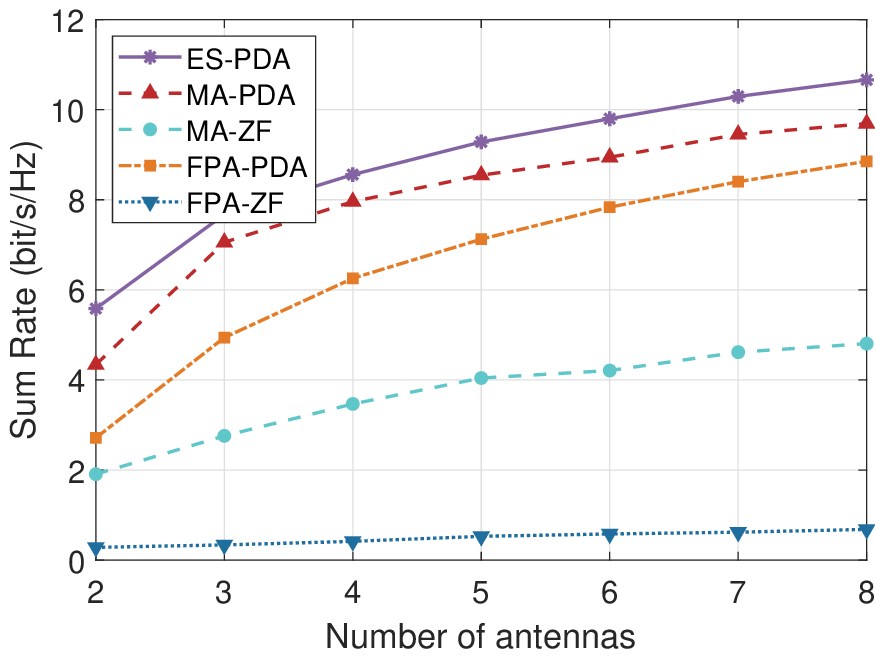}}
	\hfil
	\subfloat[Sum rate versus maximum AoD error.\label{fig:figure3}]{\includegraphics[width= 4 cm]{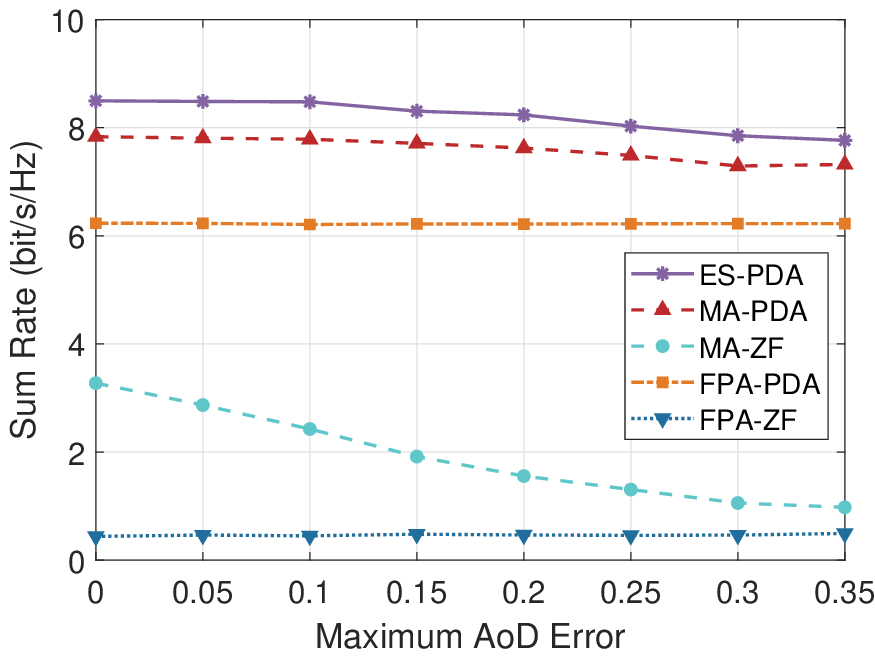}}%
	\hfil
	\subfloat[Sum rate versus normalized region size.\label{fig:figure4}]{\includegraphics[width= 4 cm]{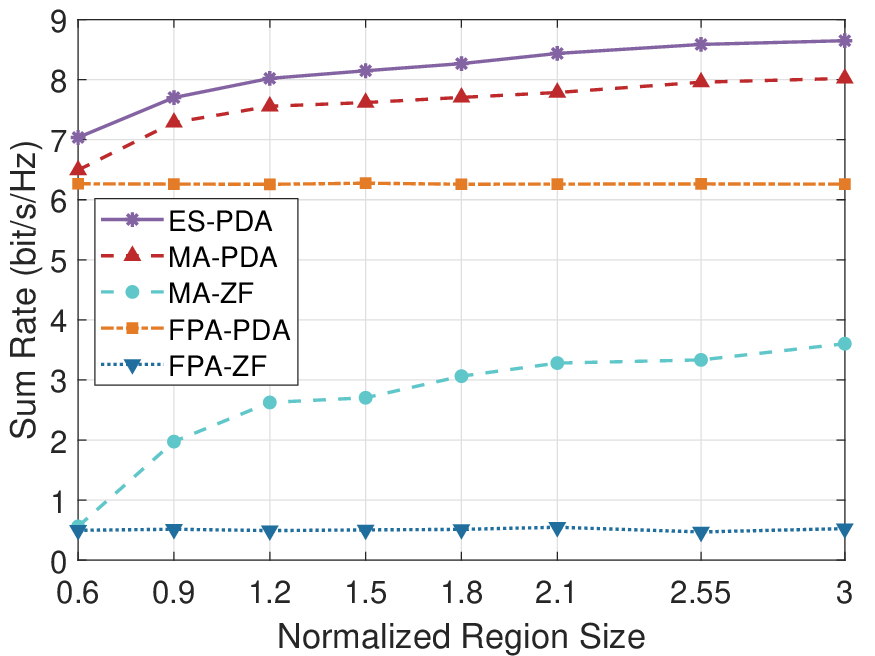}}%
	\caption{Performance comparison of different schemes in the considered covert communication system.}
	\label{fig:simu}
	\vspace{-2em}
\end{figure*}
In this section, we provide numerical results to evaluate the performance of the proposed MA solution for enhancing covert communications. In the simulations, we set $K=2$, $N=4$, and $L_k=4,1 \le k \le K$. The size of the moving region is set as $A = 3\lambda$ with $\lambda = 0.1$ m. We set the allowed minimum distance between any two MAs as $d_{\rm min} = \frac{\lambda}{2}$. The penalty factor is initialized as $\varrho = 0.05$, and $\kappa$ is set to be $1.1$. The noise power at the users is set to be $\sigma^2 = -90$ dBm and the nominal noise power at the warden is also set as $\hat{\sigma}_w^2 = -90$ dBm. In addition, $\tau$ and $\varepsilon$ are set to be $1.5$ and $0.05$, respectively. We refer \cite{zhu2024modeling} to set the other parameters. For performance comparison, we define four benchmark methods, namely ``ES-PDA", ``MA-ZF", ``FPA-PDA", and ``FPA-ZF", respectively, while our proposed algorithm is labeled as ``MA-PDA". In particular, ``ES" performs an exhaustive search for each antenna position over a 2D grid with $0.015\lambda$  spacing, which can yield a performance upper bound for the SCA-based solution. ``PDA" indicates that Algorithm \ref{alg1} is employed for transmit beamforming. ``ZF" means that zero-forcing method is utilized for the beamformer design, where the covertness constraint is satisfied by reducing the transmit power.

As observed in Fig. \ref{fig:simu}, the proposed solution is superior to all the benchmarks in terms of the sum rate and approaches the performance of the ``ES-PDA" method. It is noted that ``ES-PDA" method incurs higher computational complexity for higher performance gain. In addition, MA-enabled covert communications achieve higher sum rate than that of traditional FPAs, which demonstrates that MAs can fully exploit the spatial DoF by adjustment of antenna positions to enhance the desired signal, reduce mutual interference between users, and decrease power leakage to the warden. As shown in Fig. \ref{fig:simu}\subref{fig:figure1}, with the increase of the maximum transmit power, the sum rate increases and the performance gap between ``PDA" and ``ZF" becomes large, which demonstrates the significance of the transmit beamforming when considering the covertness constraint. It is depicted that the increasing speed of the sum rate becomes small. The reason is that the transmit power is limited by the covertness constraint, thus hindering the further increase of the sum rate. From Fig. \ref{fig:simu}\subref{fig:figure2}, we can see that the sum rates for all schemes increase due to the increasing spatial DoF brought by more antennas. In Fig. \ref{fig:simu}\subref{fig:figure3}, we evaluate the influence of imperfect field-response information on the MA positioning optimization similar to \cite{xiao2023multiuser}. It is shown that MA-enabled covert communication schemes still achieve higher sum rates compared to FPA-based schemes even for large values of the maximum AoD errors. As shown in Fig. \ref{fig:simu}\subref{fig:figure4}, the sum covert rate achieved by the MA scheme increases with the moving region size. This is because a larger region size provides more DoFs for optimization of MAs' positions. It is also noted that the sum covert rate tends to be convergent as the region size increases, which is because the spatial variation of wireless channels has been fully exploited due to their inherent periodicity in the spatial domain \cite{zhu2024modeling,zhu2024movable}. Thus, in practice, a small antenna moving region of several-wavelength size suffices to achieve high performance gains for MA-aided covert communication systems.
\vspace{-0.4cm}
\section{Conclusion}
In this letter, we investigated MA-enhanced multiuser covert communications with noise uncertainty. We formulated an optimization problem to maximize the sum covert rate by jointly optimizing the transmit beamforming and the positions of MAs at the AP. To address this highly non-convex optimization problem, a BSUM-based algorithm was proposed, where a PDA-based method was employed to tackle the transmit beamforming sub-problem and the SCA technique was utilized to solve the MA position optimization sub-problem. Simulation results illustrated that by benefiting from the new DoF in antenna position optimization, the proposed MA-enhanced covert communication system achieves superior performance in terms of sum rate compared to conventional systems with FPAs.
\vspace{-0.5cm}

\linespread{0.83}
\bibliographystyle{IEEEtran} 
\bibliography{Movable_Antennas_Enhanced_Covert_Communications}

\vfill

\end{document}